\documentclass{article}

\usepackage{arxiv}

\usepackage[utf8]{inputenc} 
\usepackage[T1]{fontenc}    
\usepackage{hyperref}       
\usepackage{url}            
\usepackage{booktabs}       
\usepackage{amsfonts}       
\usepackage{nicefrac}       
\usepackage{microtype}      
\usepackage{lipsum}
\usepackage{amsmath, amssymb}
\usepackage{comment}
\usepackage{multirow}
\usepackage{graphicx} 
\newcommand{\Return}{R}

\title{A Robust Transferable Deep Learning Framework for Cross-sectional Investment Strategy}

\author{
  Kei Nakagawa\\
  Innovation Lab\\
  Nomura Asset Management Co.,\\
  1-11-1 Nihonbashi, Chuo-ku, Tokyo, 103-8260, Japan \\
  \texttt{kei.nak.0315@gmail.com} \\
   \And
  Masaya Abe\\
  Innovation Lab\\
  Nomura Asset Management Co.,\\
  1-11-1 Nihonbashi, Chuo-ku, Tokyo, 103-8260, Japan \\
  \texttt{masaya.abe.428@gmail.com} \\
   \AND
  Junpei Komiyama\\
  Leonard N. Stern School of Business \\
  New York University \\
  44 West 4th Street, New York, NY 10012 \\
  \texttt{junpeikomiyama@gmail.com} \\
}

\begin{document}
\maketitle

\begin{abstract}
Stock return predictability is an important research theme as it reflects our economic and social organization, and significant efforts are made to explain the dynamism therein. 
Statistics of strong explanative power, called ``factor'' \cite{fama1992} have been proposed to summarize the essence of predictive stock returns.
Although machine learning methods are increasingly popular in stock return prediction \cite{mlforstock2019}, an inference of the stock returns is highly elusive, and still most investors, if partly, rely on their intuition to build a better decision making.
The challenge here is to make an investment strategy that is consistent over a reasonably long period, with the minimum human decision on the entire process.
To this end, we propose a new stock return prediction framework that we call Ranked Information Coefficient Neural Network (RIC-NN). 
RIC-NN is a deep learning approach and includes the following three novel ideas: (1) nonlinear multi-factor approach, (2) stopping criteria with ranked information coefficient (rank IC), and (3) deep transfer learning among multiple regions. 
Experimental comparison with the stocks in the Morgan Stanley Capital International (MSCI) indices shows that RIC-NN outperforms not only off-the-shelf machine learning methods but also the average return of major equity investment funds in the last fourteen years.
\end{abstract}


\section{Introduction}
Stock return predictability has been an important research theme as it reflects our economic and social organization.
Although the dynamic nature of our economic activity makes it harder to predict the future returns of the stocks, significant efforts are made to explain the dynamism therein. 
Statistics of strong explanative powers, called ``factor'' \cite{fama1992}, are proposed to summarize the essence of predictive stock returns, and a large portion of investors develop their portfolio strategies based on these factors.
For example, Book-value to Price ratio (net asset of a company divided by the market value of the corresponding stock) is one of the nominal factors, and this factor combined with simple sorting portfolio yields a positive return \cite{fama1992}. 
Due to their predictive power and robustness, investment decisions by professional investors are heavily dependent on the factors.

Machine learning is an increasingly popular tool for predicting unknown target variables; the last decades saw many attempts to apply machine learning algorithms to support smart decision-making in different financial segments \cite{atsalakis2009,cavalcante2016,mlforstock2019}.
Still, its highly elusive nature makes it harder to make a consistent inference: Most investors, if partly, rely on their intuition to build a better decision-making.

The challenge in this paper is to make an investment strategy that is consistent over a fairly long period
, with the smallest human intervention on the entire process. 
We propose a novel approach, called Rank Information Coefficient Neural Net (RIC-NN) for developing an investment strategy.  
Most of the quantitative investment strategies require a ranking over the stock returns, and we made a ranking by using a deep learning (DL) approach. 
In particular, the largest advantage of the machine learning lies in its capability to learn the nonlinear relationship between the factors and the stock returns \cite{levin1996,fan2001}, and the DL approach is recently reported to outperform other more traditional approaches in many domains, such as natural language processing \cite{DBLP:conf/nips/SutskeverVL14}, media recognition \cite{DBLP:journals/nn/CiresanMMS12}, and time-series forecasting. 

\begin{figure}
  \centering
  \includegraphics[width=0.8\hsize]{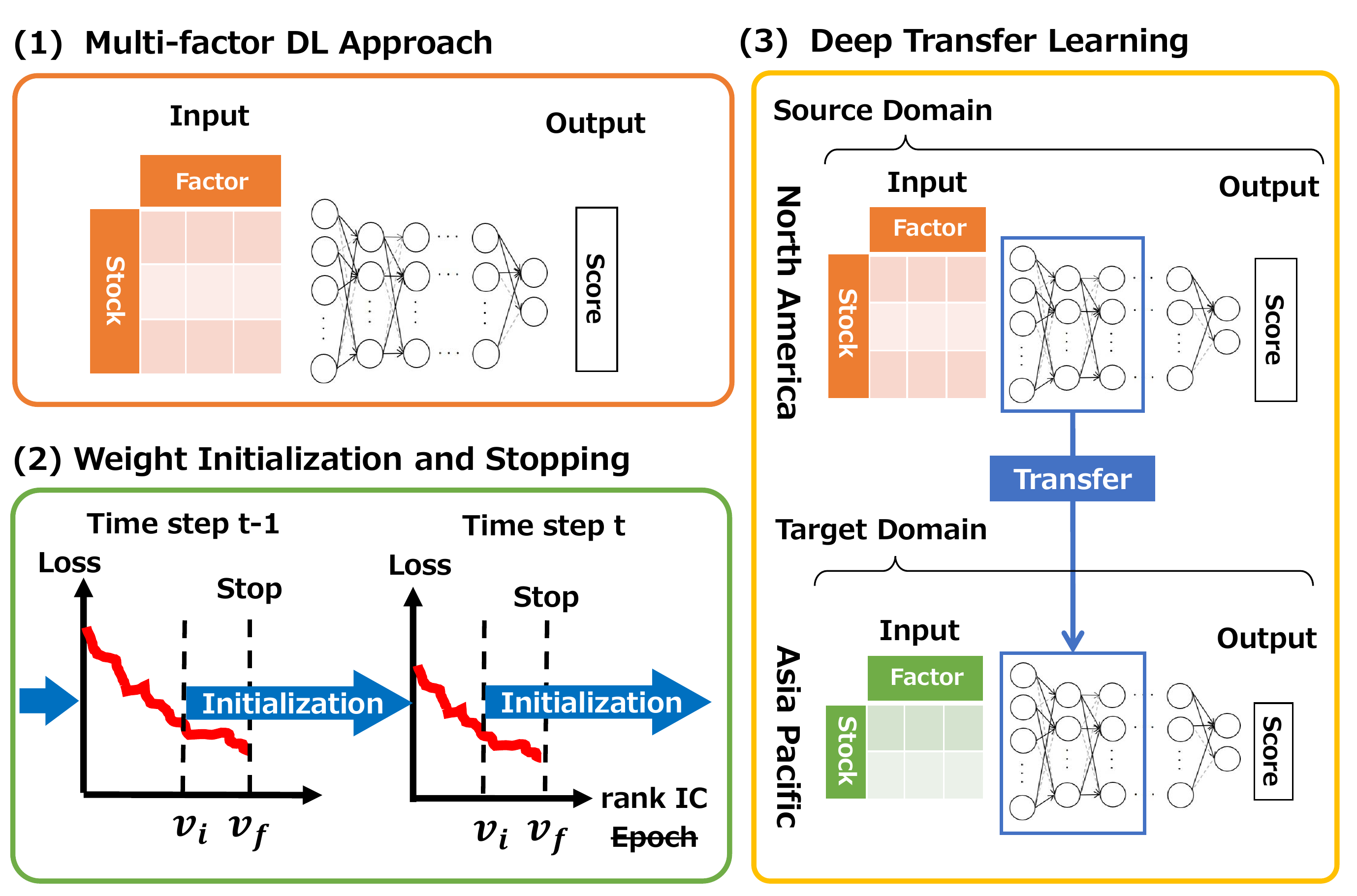}
  \caption{Our approach: RIC-NN.}
  \label{fig:framework}
\end{figure}

Due to the dynamic nature of our economic activity, naive use of the off-the-shelf machine learning tool easily overfits the existing data, and thus it fails to predict the future stock returns.
For example, \cite{chong2017} applied deep learning to stock market prediction: They reported that the advantage of a deep learning model over a linear autoregressive model has mostly disappeared in the test set. 
We show that the proposed RIC-NN consistently outperforms other methods based on off-the-shelf machine learning algorithms. 
Our framework involves three novel ideas (Figure \ref{fig:framework}): Namely, (1) we propose a deep learning multi-factor approach that enables cross-sectional prediction, and (2) the approach involves a novel training method of neural network based on the rank IC. 
Our framework is very practical: 
We conducted a comprehensive evaluation of our approach based on the stocks in the Morgan Stanley Capital International (MSCI) indices. 
Our evaluation demonstrated that a neural network with a standard training method performs poorly, whereas our RIC-NN alleviates overfitting and outperformed linear models and ensemble-based models. 
Moreover, the average return of RIC-NN over fourteen years surpasses the ones of major equity investment funds.
(3) We furthermore considered an information aggregation among several different markets in MSCI indices: Namely, a transfer learning between the North America (NA) region and the Asia Pacific (AP) region.
The experimental results imply that one can utilize the NA data to predict the future returns of the AP market, but not vice versa. 
The results verify the asymmetric causal structure between the two markets \cite{cheung1996,rejeb2016}.

\section{Related Work} 
There are two major strategies in stock trading: 
Namely, the one based on time-series analysis \cite{cavalcante2016} and the one based on cross-sectional analysis \cite{subrahmanyam2010}.

The methods of the former strategy analyze past stock prices as time-series data \cite{atsalakis2009} and are applied to a practical trading strategy that focuses on a particular stock. 
Indeed, financial time-series forecasting can be considered one of the significant challenges in time series and machine learning literature \cite{tay2001}. 
The study of financial time-series was originally started from a linear model, such as the autoregressive (AR) model in which the parameters are uniquely determined \cite{hamilton1994}. 
Introduction of machine learning techniques to the literature enabled us to capture nonlinear relationship among relevant factors without prior knowledge about the input data distribution \cite{atsalakis2009}. Still, an application of nonlinear methods to time-series data is highly non-trivial \cite{chong2017}.

The methods of the latter strategy, which include the work in this paper, perform a regression analysis using cross-sectional data of corporate attributes. 
Such a strategy aims to build a portfolio for investing as a subset of a large bucket of stocks and is applied to a practical quantitative investment strategy \cite{grinold2000}.
One of the most significant interests in a cross-sectional analysis lies in finding ``factors'' that have strong predictive powers to the expected return of a cross-sectional trading strategy:
The Fama-French three-factor model \cite{fama1992,fama1993} is one of the nominal works in this field.
They argued that the cross-sectional structure of the stock price can be explained by three factors: Namely, the beta (market portfolio), the size (market capitalization), and the value (Book-value to price ratio; BPR).
This argument inspires many subsequent research papers that propose more sophisticated versions of factors. 
\cite{harvey2016} surveyed the history of the proposed factors and argued that the number of reported factors shows a rapid increase in the last two decades:
Each year from 1980 to 1991 we saw a new factor, whereas each year from 1991 to 2003 we saw five new factors. 
In the period from 2003 to 2012, the number of factors proposed at each year rose sharply to around 18. 
As a result, over 300 factors were discovered until 2012.  

\begin{figure}
 \begin{minipage}{\hsize}
  \begin{center}
   \includegraphics[width=0.8\hsize]{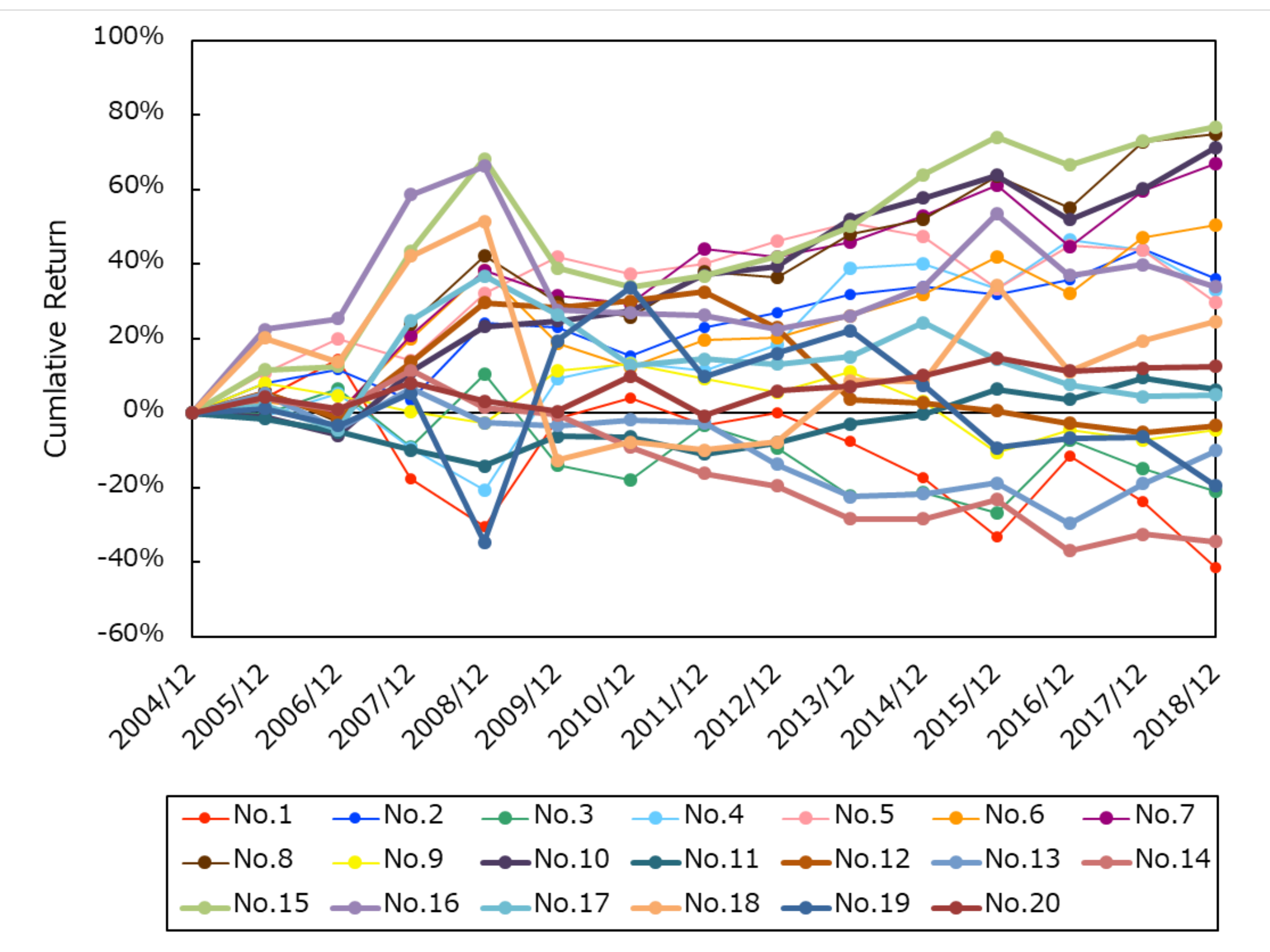}
  \end{center}
  \caption{Cumulative portfolio returns in MSCI North America based on each single factor. Factors are listed in Appendix A.}
  \label{fig:NAReturn}
 \end{minipage}
 \begin{minipage}{\hsize}
  \begin{center}
   \includegraphics[width=0.8\hsize]{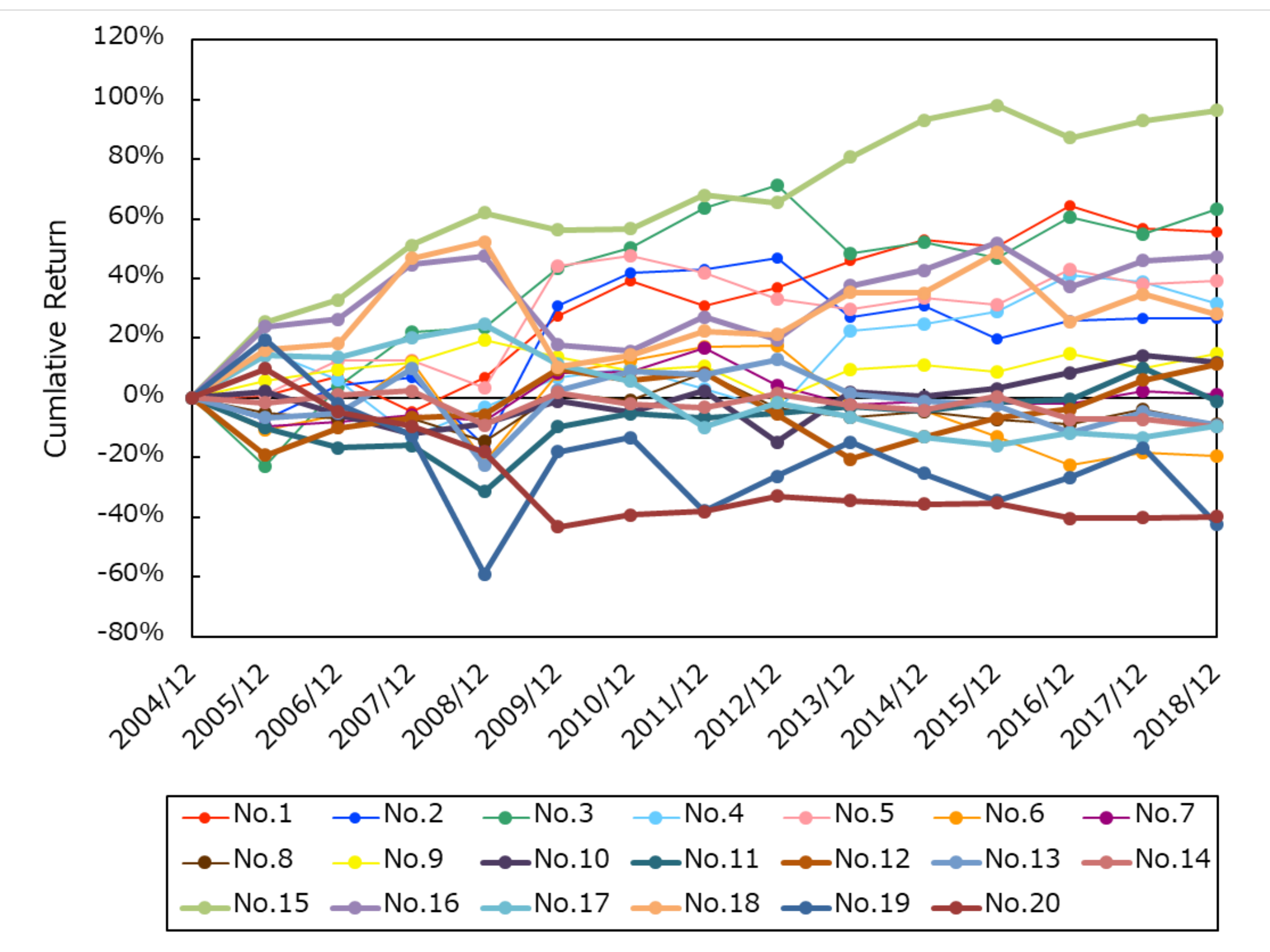}
  \end{center}
  \caption{Cumulative portfolio returns in MSCI Pacific based on each single factor.}
  \label{fig:PFReturn}
 \end{minipage}
\end{figure}

While each factor shows a positive correlation to the investment strategy, the effectiveness of the factors significantly varies over time and among different markets:
Figure \ref{fig:NAReturn} and \ref{fig:PFReturn} show the cumulative returns of the long-short portfolio strategy based on each single factor in Appendix A , from which one can see that the return of each single factor varies largely over time.
These results motivate a multi-factor approach where more than one factor is taken into consideration to the aim of better returns \cite{levin1996,fan2001}. 

In particular, machine learning approaches, which can capture the nonlinear relationship among multiple factors, are recently applied to a cross-sectional analysis.
\cite{chinco2019} applied the LASSO \cite{tibshirani1996} in the U.S. stock market, \cite{deepportfolio2016} applied an auto-encoder based nonlinear model into a U.S. biotechnology market, and \cite{abe2018,nakagawa2018,sugitomo2018} applied deep learning in the Japanese stock market.
However, these results are not universal: their experiments are performed only in a single market. Note also that, the neural nets by \cite{abe2018,nakagawa2018,sugitomo2018} adopted epoch-based stopping, which we show in the Experiments Section to be sensitive to the number of epochs.

\section{Method}
\label{sec_method}

This section describes RIC-NN, a deep learning based investment strategy.

\subsection{Cross-sectional Investment}
\label{subsec_cst}
We consider a medium-term investment cycle, where an investment is done on a monthly basis.
Namely, let $t = 1,\dots,T$ be the time step, and each step corresponds to the end of a month between December 1994 and December 2018.
We use the term ``stock universe'' (or simply universe) $U_t$ to represent all the stocks of interest at time step $t$: In the case of the North America stock market, the number of stocks in the each $U_t$ is about 700.
Note that $U_t$ gradually changes over the time step to reflect economic activities among different sectors.
At each time step, let $i \in U_t$ be an index denoting each stock in the universe. 
Let $\Return_{i,t} \in \mathbb{R}$ be the (unit) return of the stock $i$ between the time step $t-1$ and $t$.
Let $\mathbf{x}_{i,t} \in \mathbb{R}^{20}$ be the $20$ factors associated with the stock $i$ at $t$. 

In this paper, we consider investment strategies that are widely used in the literature of finance \cite{fama1992,mclean2016}.
Namely, (i) the long portfolio strategy, and (ii) the long-short portfolio strategy. 
We consider an equally-weighted (EW) portfolio, which is simple yet sometimes outperforms more sophisticated alternatives \cite{demiguel2009}. 
(i) The long portfolio strategy considered here buys the top quintile (i.e., one-fifth) of the stocks with equal weight aiming to outperform the average return of all the stocks. Namely, let $L_t \subset U_t: |L_t| = 1/5 |U_t|$ be the long portfolio. The return from the portfolio is defined as the average return of $L_t$.
\[
  \Return^L_t = \frac{1}{|L_t|} \sum_{i \in L_t} R_{i,t}
\]
(ii) The long-short portfolio strategy not only buys the top quintile of the stocks but also sells the bottom quintile of the stocks. Namely, let $S_t \subset U_t: |S_t| = 1/5 |U_t|$ be the short portfolio. Let $\Return^S_t = \frac{1}{|S_t|} \sum_{i \in S_t} \Return_{i,t}$. The average return in this strategy is defined as $\Return^{LS}_t = \Return^L_t - \Return^S_t$.
While the long-short portfolio cannot take advantage of the stock market growth, it is robust against a large market crisis (i.e., the financial crisis during 2007-2008) because of its neutral position.

Essentially, both of the strategies above requires a ranking over the expected return of the stocks in the universe since we invent on the most promising stocks. Namely, let $\mathbf{o}_t \in \mathbb{N}^{|U_t|}$ be the ground-truth ranking with its element $o_{i,t} \in \{1,2,\dots,|U_t|\}$ denotes the corresponding place for each $i \in U_t$. At each round $t$, we build the estimated ranking $\hat{\mathbf{o}}_t$. We choose $L_t$ and $U_t$ be the top and bottom quintile on the basis of $\hat{\mathbf{o}}_t$, respectively. 
Ranked information coefficient (rank IC), which is also referred as the Spearman's correlation coefficient, between two rankings $\mathbf{o}_t, \hat{\mathbf{o}}_t$ is defined as
\[
\mathrm{rank\hspace{1mm}IC}(\mathbf{o}_t, \hat{\mathbf{o}}_t) = 1 - \frac{6 \sum_{i \in U_t} (o_{i,t} - \hat{o}_{i,t})^2 }{|U_t| (|U_t|^2 - 1)},
\]
which takes the value in $[-1,1]$ and is widely used in the field of finance \cite{grinold2000}.
The larger the value of the rank IC is, the better a portfolio strategy based on the ranking is. 

We consider a rolling-horizon setting: Namely, at each time step $t$, we estimate the ranking of the next time step $\hat{\mathbf{o}}_{t+1}$. The following sections introduce RIC-NN, our DL-based method to build $\hat{\mathbf{o}}_{t+1}$.

\subsection{Feature Augmentation}

The normalized rank of the stock $i$ at time step $t$ is denoted as $r_{i,t} \in \mathbb{R}$: 
Namely, we rank the stocks in accordance with their return $\{\Return_{i,t}\}$ and normalize them so that $r_{i,t} \in [0,1]$ (i.e., $r_{i,t}$ for the stock of the largest return at each $t$ is $1$, whereas $r_{i,t}$ for the stock of the median return is $0.5$).

At each time step $t$, we build an estimator $\hat{r}_{i,t}$ of $r_{i,t}$ by using the following augmented feature vector $\mathbf{v}_{i,t} \in \mathbb{R}^{180}$: Namely, given that many of the factors are updated in quarterly basis (i.e., each $3$ time steps), we define 
$\mathbf{v}_{i,t}=(\mathbf{x}_{i,t},\mathbf{x}_{i,t-3},...,\mathbf{x}_{i,t-12},\mathbf{x}_{i,t} /^R \mathbf{x}_{i,t-3},...,\mathbf{x}_{i,t} /^R \mathbf{x}_{i,t-12}) \in \mathbb{R}^{180}$ using the past five time steps, where $\mathbf{x}/^R\mathbf{y}$ over two vectors $\mathbf{x}$ and $\mathbf{y}$ denotes an element-wise differentiation operator with its each element is defined by $2 \times (x-y)/(|x|+|y|)$, which is popularly used in finance \cite{rosenberg1973}.

\subsection{Prediction Model: Neural Net Architecture}

We adopt a seven-layer feed forward neural network with Rectified linear function (ReLU) activation function \cite{Hahnloser2000DigitalSA} to learn the relationship between $v_{i,t}$ and $r_{i,t+1}$. 
The hidden layer size is set to be $(150-150-100-100-50-50)$, and the dropout rate for each layer is set to be $(50\%-50\%-30\%-30\%-10\%-10\%)$. 

\subsection{Loss Function and Optimization}

We adopt the standard mean squared error (MSE) as the loss function and train our deep learning model by using the data of the latest $120$ time steps from the past 10 years. Namely, 
\begin{equation}
{\bf MSE}_{t} = \frac{1}{K}\left\{\sum_{t'=t-N}^{t-1} \sum_{i \in U_{t'}} (r_{i,t'+1}-f(\mathbf{v}_{i,t'}; \boldsymbol{\theta}_{t'}))^2 \right\},
\label{MSE}    
\end{equation}
where $N = 120$ (i.e., ten years) is the size of sliding window to consider and $K = \sum_{t'=t-N}^{t-1} |U_t'|$ is the number of all training examples, and $f(\cdot, \boldsymbol{\theta})$ is our neural net with weight parameter $\boldsymbol{\theta}$.
We adopt the Adam \cite{kingma2014} optimizer and batch normalization \cite{ioffe2015}. 
The mini-batch size is set to be $300$.  

\subsection{Initialization and Stopping Criteria}
\label{subsec_stopping}
A fundamental challenge in the cross-sectional analysis lies in its dynamism: 
Standard machine learning methods focus on the generalization performance on the i.i.d. assumption where the training dataset and the test dataset are drawn from the same (unknown) underlying distribution. 
However, a straightforward application of deep neural networks leads to overfitting to the current time window, which compromises the performance as a predictor of the next time step.

To avoid overfitting, we initialize and terminate the training in the following criterion ((2) in Figure \ref{fig:framework}): Namely, we define the initialization rank IC $v_i \in [0,1]$ and stopping rank IC $v_s \in [0,1]$, and conducts the training as follows. Let $\boldsymbol{\theta}_{t, v}$ is the weights of the RIC-NN at time step $t$ during the training when the average from rank IC in the training window reaches $v$. We used (i) $\boldsymbol{\theta}_{t-1, v_i}$ as the initial parameters to train model at time step $t$ and (ii) adopts $\boldsymbol{\theta}_{t, v_s}$ as the final model parameter $\boldsymbol{\theta}_{t}$. We estimate  $\hat{\mathbf{o}}_{t+1}$ by ranking the stocks in accordance with $f(\mathbf{v}_{i,t}; \boldsymbol{\theta}_{t})$, which, combined with the long or the long-short portfolio, defines our RIC-NN.

The value of $v_i, v_f$ are set to be $0.16, 0.20$. These values are optimized by the performance in the three years from 2005 to 2008. Essentially, these values ``moderately overfit'' to the model: The value of the rank IC of a fairly good portfolio is around $0.1$, and thus these values are large enough to exploit current data while it stops before overfitting to the dataset in the current window.
The experiment section shows that these values consistently performs well in multiple markets of very different natures.

\subsection{Performance Measures}
\label{subsec_measures}

In evaluating an investment strategy, we use the following measures that are widely used in the field of finance \cite{brandt2010}. These measures evaluate not only the actual return of the portfolio but also the magnitude of the risk taken: A sample from highly fluctuated series can cause a large variance, and thus a return normalized by a risk yields a more reliable evaluation.

Regarding the long portfolio strategy, the annualized return is the excess return (Alpha) against the average return of all stocks in the universe, the risk (tracking error; TE) is calculated as the standard deviation of Alpha and risk/return is Alpha/TE (information ratio; IR).
\begin{align}
    {\bf Alpha} &= \prod_{t=1}^T (1+\alpha_{t})^{12/T}-1 \label{eq_alpha}\\
    {\bf TE} &= \sqrt{\frac{12}{T-1}\times(\alpha_{t}-\mu_{\alpha})^2}\\
    {\bf IR} &= {\bf Alpha}/{\bf TE}
\end{align}

Here, $\alpha_{t} =  \Return_t^L - \frac{1}{|U_t|}\sum_{i \in U_t} \Return_{i,t}$ , $\mu_{\alpha}= (1/T) \sum_{t=1}^T \alpha_{t}$.

Likewise, we evaluate the long-short portfolio strategy by its annualized return (AR), risk as the standard deviation of return (RISK), risk/return (R/R) as return divided by risk as for the long portfolio strategy.
\begin{align}
    {\bf AR} &= \prod_{t=1}^T (1+\Return_t^{LS})^{12/T}-1 \\
    {\bf RISK} &= \sqrt{\frac{12}{T-1}\times(\Return_t^{LS}-\mu^{LS})^2}\\
    {\bf R/R} &= {\bf AR}/{\bf RISK}
\end{align}

Here, $\mu^{LS}= (1/T) \sum_{t=1}^T \Return_t^{LS}$ be the average return of the long-short portfolio.

In summary, the return of the long (resp. long-short) portfolio is evaluated by Alpha (resp. AR), whereas the risk of the long (resp. long-short) portfolio is evaluated by TE (resp. RISK). We use the risk-normalized return (i.e. IR for the long and R/R for the long-short) that gives more reliable measure than the return itself.

We also evaluate maximum drawdown (MaxDD), which is yet another widely used risk measures \cite{magdon04,Shen:2017:PSV:3298239.3298460}, for both of the long portfolio strategy and the long-short portfolio strategy:
Namely, MaxDD is defined as the largest drop from an extremum:
\begin{align}
    {\bf MaxDD} &= \min_{k \in [1,T]}\left(0,\frac{W_k^{\mathrm{Port} }}{\max_{j \in [1,k]} W_j^{\mathrm{Port}}}-1\right) \\
    W_k^{\mathrm{Port}} &= \prod_{i=1}^k (1+\Return_i^{\mathrm{Port}}).
\end{align}
where $\Return_i^{\mathrm{Port}} = \Return_i^{L}$ (resp. $\Return_i^{\mathrm{Port}} = \Return_i^{LS}$) for the long (resp. long-short) strategy.
These performance measures are calculated monthly during the prediction period from January 2005 to December 2018 ($T=168$).

\section{Experiments}
\label{sec_exp}

\subsection{Dataset}

We prepare a stock dataset corresponding to Morgan Stanley Capital International (MSCI) North America and MSCI Pacific Indices. 
These MSCI indices comprise the large and mid-cap segments of the North America (NA) and Asia Pacific (AP) markets respectively, and are widely used as a benchmark for the institutional investors investing in each stock market \cite{chen2019}. 
We use $20$ popular factors listed in Appendix A.
In calculating these factors, we use the following data sources: Namely, Compustat, WorldScope, Thomson Reuters, the Institutional Brokers' Estimate System (I/B/E/S), and EXSHARE. 
Combining these sources, we calculate the factors on a monthly basis.
As for stock returns, local returns with dividends are acquired.

By using these sources, we build a dataset comprised 1,194 stocks on average (NA: 702, AP: 492) and 288 time steps from December 1994 to November 2018. 
The following sections show the performance of the proposed RIC-NN strategy compared with several baselines. 
This dataset involves a reasonably long period so that we can evaluate a consistently-good investment strategy.

\subsection{Comparison with Off-the-Shelf Models}

We compare the performance of RIC-NN with major off-the-shelf machine learning algorithms. 
Namely: LASSO regression (LASSO) model \cite{tibshirani1996}, random forest (RF), and standard Neural Network (NN). LASSO and RF are implemented with scikit-learn \cite{pedregosa2011}, and NN is implemented with TensorFlow \cite{abadi2016}. These methods are used to learn the relation between $v_{i,t}$ and $r_{i,t+1}$. Regarding the hyperparameters, regularization strength ("alpha") of LASSO is set to 0.001, which is the largest value to yield a meaningful ranking. We use the default hyperparameters of RF. 
Several different hyperparameters are tested, and their results are shown in Appendix B.
NN adopted the same framework as our RIC-NN, except for the fact that NN stops the training at Epoch $56$ in MSCI North America and $46$ in MSCI Pacific\footnote{These epochs are chosen so that the rank IC reaches 0.20 during the training of the first time step.}. We used random numbers as initial weights for the first time step.

Table \ref{tbl:NA} compares the algorithms in the MSCI NA dataset. 
RIC-NN outperforms all of the LASSO, RF, and NN in both of the risk and the return measures, regardless of whether the portfolio strategy is the long or the long-short. 
A notable finding is that RF and NN have smaller returns compared with LASSO. 
Our hypothesis is that the highly non-stationary nature of the stocks has lead to the overfitting of these nonlinear models. 
Table \ref{tbl:PF} shows the results in the MSCI Pacific dataset. Although LASSO yields a larger return than RIC-NN by taking a larger risk, in terms of a risk-normalized return, which is the prominent measure of investment strategy, RIC-NN outperforms the other methods. Regarding the results of the transfer learning (``TF from AP/NA''), we discuss in a later section. 

We have also conducted the same experiment with Ridge Regression (RR): The performance of RR is not very different from that of LASSO.

\begin{table*}[]
\centering
\caption{Experimental Results of Long portfolio and Long-Short portfolio in MSCI North America. Bold characters indicate the best ones among each category. The evaluation measures are the ones discussed in the ``Performance Measures'' Section: Alpha (resp. AR) measures return, TE (resp. RISK) and MaxDD measure risk, and IR (resp. R/R) is a risk-normalized return measure in the long (resp. long-short) portfolio.}
\label{tbl:NA}
\begin{tabular}{crrrrr}
\hline
\multicolumn{1}{|c|}{\multirow{2}{*}{Long}}       & \multicolumn{1}{c|}{Linear}   & \multicolumn{4}{c|}{Nonlinear}                                                                                                                                                               \\ \cline{2-6} 
\multicolumn{1}{|c|}{}                            & \multicolumn{1}{c|}{LASSO}       & \multicolumn{1}{c|}{RF}       & \multicolumn{1}{c|}{NN} & \multicolumn{1}{c|}{RIC-NN}            & \multicolumn{1}{c|}{\begin{tabular}[c]{@{}c@{}}RIC-NN\\ (TF from AP)\end{tabular}} \\ \hline
\multicolumn{1}{|c|}{Alpha}                       & \multicolumn{1}{r|}{0.62\%}   & \multicolumn{1}{r|}{0.79\%}   & \multicolumn{1}{r|}{0.82\%}    & \multicolumn{1}{r|}{\textbf{1.23\%}}   & \multicolumn{1}{r|}{1.20\%}                                                        \\ \hline
\multicolumn{1}{|c|}{TE}                          & \multicolumn{1}{r|}{5.40\%}   & \multicolumn{1}{r|}{5.14\%}   & \multicolumn{1}{r|}{4.48\%}    & \multicolumn{1}{r|}{\textbf{4.14\%}}   & \multicolumn{1}{r|}{4.43\%}                                                        \\ \hline
\multicolumn{1}{|c|}{IR}                          & \multicolumn{1}{r|}{0.11}     & \multicolumn{1}{r|}{0.13}     & \multicolumn{1}{r|}{0.18}      & \multicolumn{1}{r|}{\textbf{0.30}}     & \multicolumn{1}{r|}{0.27}                                                          \\ \hline
\multicolumn{1}{|c|}{MaxDD}                       & \multicolumn{1}{r|}{-21.84\%} & \multicolumn{1}{r|}{-24.57\%} & \multicolumn{1}{r|}{-17.41\%}  & \multicolumn{1}{r|}{\textbf{-14.37\%}} & \multicolumn{1}{r|}{-20.57\%}                                                      \\ \hline \hline
\multicolumn{1}{|c|}{\multirow{2}{*}{Long-Short}} & \multicolumn{1}{c|}{Linear}   & \multicolumn{4}{c|}{Nonlinear}                                                                                                                                                               \\ \cline{2-6} 
\multicolumn{1}{|c|}{}                            & \multicolumn{1}{c|}{LASSO}       & \multicolumn{1}{c|}{RF}       & \multicolumn{1}{c|}{NN} & \multicolumn{1}{c|}{RIC-NN}            & \multicolumn{1}{c|}{\begin{tabular}[c]{@{}c@{}}RIC-NN\\ (TF from AP)\end{tabular}} \\ \hline
\multicolumn{1}{|c|}{AR}                      & \multicolumn{1}{r|}{2.24\%}   & \multicolumn{1}{r|}{1.71\%}   & \multicolumn{1}{r|}{2.10\%}    & \multicolumn{1}{r|}{\textbf{3.86\%}}   & \multicolumn{1}{r|}{2.16\%}                                                        \\ \hline
\multicolumn{1}{|c|}{RISK}                        & \multicolumn{1}{r|}{10.90\%}  & \multicolumn{1}{r|}{11.51\%}   & \multicolumn{1}{r|}{9.47\%}    & \multicolumn{1}{r|}{\textbf{7.85\%}}   & \multicolumn{1}{r|}{9.52\%}                                                        \\ \hline
\multicolumn{1}{|c|}{R/R}                 & \multicolumn{1}{r|}{0.21}     & \multicolumn{1}{r|}{0.15}     & \multicolumn{1}{r|}{0.22}      & \multicolumn{1}{r|}{\textbf{0.49}}     & \multicolumn{1}{r|}{0.23}                                                          \\ \hline
\multicolumn{1}{|c|}{MaxDD}                       & \multicolumn{1}{r|}{-34.73\%} & \multicolumn{1}{r|}{-42.21\%} & \multicolumn{1}{r|}{-34.49\%}  & \multicolumn{1}{r|}{\textbf{-21.26\%}} & \multicolumn{1}{r|}{-39.35\%}                                                      \\ \hline
\end{tabular}
\end{table*}

\begin{table*}[]
\centering
\caption{Experimental Results of Long portfolio and Long-Short portfolio in MSCI Pacific. Bold characters indicate the best ones among each category.}
\label{tbl:PF}
\begin{tabular}{crrrrr}
\hline
\multicolumn{1}{|c|}{\multirow{2}{*}{Long}}       & \multicolumn{1}{c|}{Linear}   & \multicolumn{4}{c|}{Nonlinear}                                                                                                                                                      \\ \cline{2-6} 
\multicolumn{1}{|c|}{}                            & \multicolumn{1}{c|}{LASSO}       & \multicolumn{1}{c|}{RF}       & \multicolumn{1}{c|}{NN} & \multicolumn{1}{c|}{RIC-NN}   & \multicolumn{1}{c|}{\begin{tabular}[c]{@{}c@{}}RIC-NN\\ (TF from NA)\end{tabular}} \\ \hline
\multicolumn{1}{|c|}{Alpha}                       & \multicolumn{1}{r|}{5.35\%}   & \multicolumn{1}{r|}{3.79\%}   & \multicolumn{1}{r|}{4.34\%}    & \multicolumn{1}{r|}{5.25\%}   & \multicolumn{1}{r|}{\textbf{5.78\%}}                                               \\ \hline
\multicolumn{1}{|c|}{TE}                          & \multicolumn{1}{r|}{5.17\%}   & \multicolumn{1}{r|}{5.75\%}   & \multicolumn{1}{r|}{4.18\%}    & \multicolumn{1}{r|}{4.20\%}   & \multicolumn{1}{r|}{\textbf{3.95\%}}                                               \\ \hline
\multicolumn{1}{|c|}{IR}                          & \multicolumn{1}{r|}{1.04}     & \multicolumn{1}{r|}{0.66}     & \multicolumn{1}{r|}{1.04}      & \multicolumn{1}{r|}{1.25}     & \multicolumn{1}{r|}{\textbf{1.46}}                                                 \\ \hline
\multicolumn{1}{|c|}{MaxDD}                       & \multicolumn{1}{r|}{-11.53\%} & \multicolumn{1}{r|}{-11.43\%}  & \multicolumn{1}{r|}{-9.37\%}   & \multicolumn{1}{r|}{-7.51\%}  & \multicolumn{1}{r|}{\textbf{-3.37\%}}                                              \\ \hline \hline
\multicolumn{1}{|c|}{\multirow{2}{*}{Long-Short}} & \multicolumn{1}{c|}{Linear}   & \multicolumn{4}{c|}{Nonlinear}                                                                                                                                                      \\ \cline{2-6} 
\multicolumn{1}{|c|}{}                            & \multicolumn{1}{c|}{LASSO}       & \multicolumn{1}{c|}{RF}       & \multicolumn{1}{c|}{NN} & \multicolumn{1}{c|}{RIC-NN}   & \multicolumn{1}{c|}{\begin{tabular}[c]{@{}c@{}}RIC-NN\\ (TF from NA)\end{tabular}} \\ \hline
\multicolumn{1}{|c|}{AR}                      & \multicolumn{1}{r|}{10.27\%}  & \multicolumn{1}{r|}{7.78\%}   & \multicolumn{1}{r|}{8.52\%}    & \multicolumn{1}{r|}{9.81\%}   & \multicolumn{1}{r|}{\textbf{10.95\%}}                                              \\ \hline
\multicolumn{1}{|c|}{RISK}                        & \multicolumn{1}{r|}{9.23\%}   & \multicolumn{1}{r|}{9.65\%}   & \multicolumn{1}{r|}{7.78\%}    & \multicolumn{1}{r|}{7.83\%}   & \multicolumn{1}{r|}{\textbf{7.14\%}}                                               \\ \hline
\multicolumn{1}{|c|}{R/R}                 & \multicolumn{1}{r|}{1.11}     & \multicolumn{1}{r|}{0.81}     & \multicolumn{1}{r|}{1.10}      & \multicolumn{1}{r|}{1.25}     & \multicolumn{1}{r|}{\textbf{1.53}}                                                 \\ \hline
\multicolumn{1}{|c|}{MaxDD}                       & \multicolumn{1}{r|}{-18.07\%} & \multicolumn{1}{r|}{-18.66\%} & \multicolumn{1}{r|}{-19.74\%}  & \multicolumn{1}{r|}{-11.06\%} & \multicolumn{1}{r|}{\textbf{-8.89\%}}                                              \\ \hline
\end{tabular}
\end{table*}

\subsection{Stopping Criteria: Rank IC versus Epoch}

Table \ref{tbl:RankICvsEpoch_NA} and \ref{tbl:RankICvsEpoch_PF} show the result of NN with different number of training epochs. 
While NN that stops at epoch $50$ performs better in the NA market, NN that stops at epoch $60$ performs better in the AP market. 
One can also find that the performance of NN is very sensitive to the choice of the stopping epoch. 
On the other hand, RIC-NN that consistently stops at $v_f=0.20$ outperforms most of (epoch-based) NN. 
This implies that the rank IC is a consistent measure of the fitness of stock prediction models.

\begin{table*}[]
\centering
\caption{Comparison between RIC-NN and NN with different number of training epochs in MSCI North America.}
\label{tbl:RankICvsEpoch_NA}
\begin{tabular}{crrrrrr}
\hline
\multicolumn{1}{|c|}{\multirow{2}{*}{Long}}       & \multicolumn{1}{c|}{\multirow{2}{*}{RIC-NN}} & \multicolumn{5}{c|}{NN (Epoch)}                                                                                                                                              \\ \cline{3-7} 
\multicolumn{1}{|c|}{}                            & \multicolumn{1}{c|}{}                        & \multicolumn{1}{c|}{40}       & \multicolumn{1}{c|}{50}       & \multicolumn{1}{c|}{56}       & \multicolumn{1}{c|}{60}              & \multicolumn{1}{c|}{80}              \\ \hline
\multicolumn{1}{|c|}{Alpha}                       & \multicolumn{1}{r|}{1.23\%}                  & \multicolumn{1}{r|}{0.18\%}   & \multicolumn{1}{r|}{\textbf{1.48\%}}   & \multicolumn{1}{r|}{0.82\%}   & \multicolumn{1}{r|}{1.25\%} & \multicolumn{1}{r|}{0.70\%}          \\ \hline
\multicolumn{1}{|c|}{TE}                          & \multicolumn{1}{r|}{\textbf{4.14\%}}         & \multicolumn{1}{r|}{4.52\%}   & \multicolumn{1}{r|}{4.35\%}   & \multicolumn{1}{r|}{4.48\%}   & \multicolumn{1}{r|}{4.49\%}          & \multicolumn{1}{r|}{4.14\%}          \\ \hline
\multicolumn{1}{|c|}{IR}                          & \multicolumn{1}{r|}{0.30}           & \multicolumn{1}{r|}{0.04}     & \multicolumn{1}{r|}{\textbf{0.34}}     & \multicolumn{1}{r|}{0.18}     & \multicolumn{1}{r|}{0.28}            & \multicolumn{1}{r|}{0.17}            \\ \hline
\multicolumn{1}{|c|}{MaxDD}                       & \multicolumn{1}{r|}{-14.37\%}       & \multicolumn{1}{r|}{-22.67\%} & \multicolumn{1}{r|}{\textbf{-13.48\%}} & \multicolumn{1}{r|}{-17.41\%} & \multicolumn{1}{r|}{-20.98\%}        & \multicolumn{1}{r|}{-15.94\%}        \\ \hline \hline
\multicolumn{1}{|c|}{\multirow{2}{*}{Long-Short}} & \multicolumn{1}{l|}{\multirow{2}{*}{RIC-NN}} & \multicolumn{5}{c|}{NN (Epoch)}                                                                                                                                              \\ \cline{3-7} 
\multicolumn{1}{|c|}{}                            & \multicolumn{1}{l|}{}                        & \multicolumn{1}{c|}{40}       & \multicolumn{1}{c|}{50}       & \multicolumn{1}{c|}{56}       & \multicolumn{1}{c|}{60}              & \multicolumn{1}{c|}{80}              \\ \hline
\multicolumn{1}{|c|}{AR}                      & \multicolumn{1}{r|}{\textbf{3.86\%}}         & \multicolumn{1}{r|}{0.67\%}   & \multicolumn{1}{r|}{3.24\%}   & \multicolumn{1}{r|}{2.10\%}   & \multicolumn{1}{r|}{2.02\%}          & \multicolumn{1}{r|}{3.10\%}          \\ \hline
\multicolumn{1}{|c|}{RISK}                        & \multicolumn{1}{r|}{7.85\%}                  & \multicolumn{1}{r|}{9.08\%}   & \multicolumn{1}{r|}{10.06\%}  & \multicolumn{1}{r|}{9.47\%}   & \multicolumn{1}{r|}{9.05\%}          & \multicolumn{1}{r|}{\textbf{7.73\%}} \\ \hline
\multicolumn{1}{|c|}{R/R}                 & \multicolumn{1}{r|}{\textbf{0.49}}           & \multicolumn{1}{r|}{0.07}     & \multicolumn{1}{r|}{0.32}     & \multicolumn{1}{r|}{0.22}     & \multicolumn{1}{r|}{0.22}            & \multicolumn{1}{r|}{0.40}            \\ \hline
\multicolumn{1}{|c|}{MaxDD}                       & \multicolumn{1}{r|}{\textbf{-21.26\%}}       & \multicolumn{1}{r|}{-40.09\%} & \multicolumn{1}{r|}{-26.20\%} & \multicolumn{1}{r|}{-34.49\%} & \multicolumn{1}{r|}{-31.62\%}        & \multicolumn{1}{r|}{-23.47\%}        \\ \hline
\end{tabular}
\end{table*}

\begin{table*}[]
\centering
\caption{Comparison between RIC-NN and NN with different number of training epochs in MSCI Pacific.}
\label{tbl:RankICvsEpoch_PF}
\begin{tabular}{crrrrrr}
\hline
\multicolumn{1}{|c|}{\multirow{2}{*}{Long}}       & \multicolumn{1}{c|}{\multirow{2}{*}{RIC-NN}} & \multicolumn{5}{c|}{NN (Epoch)}                                                                                                                                            \\ \cline{3-7} 
\multicolumn{1}{|c|}{}                            & \multicolumn{1}{c|}{}                        & \multicolumn{1}{c|}{40}       & \multicolumn{1}{c|}{46}       & \multicolumn{1}{c|}{50}       & \multicolumn{1}{c|}{60}            & \multicolumn{1}{c|}{80}              \\ \hline
\multicolumn{1}{|c|}{Alpha}                       & \multicolumn{1}{r|}{\textbf{5.25\%}}         & \multicolumn{1}{r|}{4.13\%}   & \multicolumn{1}{r|}{4.34\%}   & \multicolumn{1}{r|}{4.28\%}   & \multicolumn{1}{r|}{4.52\%}        & \multicolumn{1}{r|}{2.99\%}          \\ \hline
\multicolumn{1}{|c|}{TE}                          & \multicolumn{1}{r|}{4.20\%}                  & \multicolumn{1}{r|}{4.36\%}   & \multicolumn{1}{r|}{4.18\%}   & \multicolumn{1}{r|}{4.73\%}   & \multicolumn{1}{r|}{4.34\%}        & \multicolumn{1}{r|}{\textbf{4.06\%}} \\ \hline
\multicolumn{1}{|c|}{IR}                          & \multicolumn{1}{r|}{\textbf{1.25}}           & \multicolumn{1}{r|}{0.95}     & \multicolumn{1}{r|}{1.04}     & \multicolumn{1}{r|}{0.90}     & \multicolumn{1}{r|}{1.04}          & \multicolumn{1}{r|}{0.74}            \\ \hline
\multicolumn{1}{|c|}{MaxDD}                       & \multicolumn{1}{r|}{-7.51\%}        & \multicolumn{1}{r|}{-8.08\%}  & \multicolumn{1}{r|}{-9.37\%}  & \multicolumn{1}{r|}{\textbf{-7.16\%}}  & \multicolumn{1}{r|}{-7.45\%}       & \multicolumn{1}{r|}{-7.52\%}         \\ \hline \hline
\multicolumn{1}{|c|}{\multirow{2}{*}{Long-Short}} & \multicolumn{1}{l|}{\multirow{2}{*}{RIC-NN}} & \multicolumn{5}{c|}{NN (Epoch)}                                                                                                                                            \\ \cline{3-7} 
\multicolumn{1}{|c|}{}                            & \multicolumn{1}{l|}{}                        & \multicolumn{1}{c|}{40}       & \multicolumn{1}{c|}{46}       & \multicolumn{1}{c|}{50}       & \multicolumn{1}{c|}{60}            & \multicolumn{1}{c|}{80}              \\ \hline
\multicolumn{1}{|c|}{AR}                      & \multicolumn{1}{r|}{\textbf{9.81\%}}         & \multicolumn{1}{r|}{8.89\%}   & \multicolumn{1}{r|}{8.52\%}   & \multicolumn{1}{r|}{8.97\%}   & \multicolumn{1}{r|}{9.78\%}        & \multicolumn{1}{r|}{6.15\%}          \\ \hline
\multicolumn{1}{|c|}{RISK}                        & \multicolumn{1}{r|}{7.83\%}                  & \multicolumn{1}{r|}{7.63\%}   & \multicolumn{1}{r|}{7.78\%}   & \multicolumn{1}{r|}{8.05\%}   & \multicolumn{1}{r|}{7.73\%}        & \multicolumn{1}{r|}{\textbf{7.18\%}} \\ \hline
\multicolumn{1}{|c|}{R/R}                 & \multicolumn{1}{r|}{1.25}                    & \multicolumn{1}{r|}{1.16}     & \multicolumn{1}{r|}{1.10}     & \multicolumn{1}{r|}{1.11}     & \multicolumn{1}{r|}{\textbf{1.26}} & \multicolumn{1}{r|}{0.86}            \\ \hline
\multicolumn{1}{|c|}{MaxDD}                       & \multicolumn{1}{r|}{\textbf{-11.06\%}}       & \multicolumn{1}{r|}{-13.07\%} & \multicolumn{1}{r|}{-19.74\%} & \multicolumn{1}{r|}{-12.17\%} & \multicolumn{1}{r|}{-14.70\%}      & \multicolumn{1}{r|}{-13.44\%}        \\ \hline
\end{tabular}
\end{table*}

\subsection{Comparison between NA and AP markets}

The value of Alpha (Eq. \eqref{eq_alpha}) indicates the advantage of the long strategy over the average return in the universe, which enables us to infer the possible advantage we can obtain by using machine learning algorithms. 

Comparing the Alpha in Table \ref{tbl:NA} and \ref{tbl:PF}, machine learning algorithms has a smaller advantage in the NA market than they do in the AP market:

A portfolio strategy of a higher return essentially exploits the gap between the market value of the stocks and the true valuation of the companies: The more efficient a market is, the more difficult obtaining a higher return is. In other words, the result implies the efficiency of the NA market compared with the AP market.

\subsection{Transfer Learning}

To exploit the interdependency between the markets, we further apply transfer learning to our RIC-NN. Namely, we use the weights of the first four layers that are trained in the source region as the initial weight of the target region.

Table \ref{tbl:NA} shows that the transfer from NA to AP is not very successful, whereas \ref{tbl:PF} shows the transfer from AP to NA is quite successful. In other words, NA as a source domain is quite informative to enhance the performance of AP, not vice versa. Those results are consistent with the market movements propagate from the NA stock market to the AP stock market \cite{cheung1996,rejeb2016}.
The experiment here shows the capability of RIC-NN to exploit highly non-trivial causal structure among multiple markets by using deep neural networks. 

\subsection{Comparison with Actual Investment Funds}

This section compares the performance of RIC-NN with major funds where the investments involve decision-making by human experts.
We select the top 5 funds in terms of the total assets (US dollar) excluding index funds as following criteria and calculate average total return series of these funds, including the trust fees:
Namely, we select these funds by querying Bloomberg fund screening search with the following condition:
\begin{itemize}
    \item Fund Asset Class Focus: Equity
    \item Fund Geographical Focus: North America Region (resp. Asian Pacific Region)
    \item Fund Type: Open-End-Funds
    \item Currency: US dollar
    \item Market Cap Focus (Holdings Based): Large-cap, Mid-cap
    \item Inception Date: before 12/31/2004
\end{itemize}

In both of the NA and AP regions, the correlation coefficient between the performance of the averaged funds above and the MSCI index is larger than $0.9$, which implies these funds are based on the long strategy.
For comparison, We add the benchmark returns calculated by average return of MSCI North America (resp. MSCI Pacific) constituent to long portfolio strategy performance and convert to US dollars. 

Table \ref{vsHuman} shows the performance of RIC-NN and the aforementioned stock investing funds from January 2005 to June 2018. The corresponding time-series data is shown in the supplementary material (see Appendix C).
Unlike the performance of the machine learning models, the performance of the funds involves the transaction cost: As a conservative baseline, Table \ref{vsHuman} shows the performance of RIC-NN where the transaction cost for updating the entire portfolio is deducted (i.e., an overestimated transaction cost\footnote{We have deducted the cost of rebalancing all the stocks in the portfolio every month. Estimated transaction cost is 0.05\% one way in North America and 0.1\% one way in Asia Pacific.}): RIC-NN still outperforms the average performance of the funds.

\begin{table}[]
\caption{The upper panel: Performance of RIC-NN in MSCI North America and averaged performance of five investment funds in the NA stock market. The lower panel: Performance of RIC-NN (TF from NA) in MSCI Pacific and averaged performance of five investment funds in the AP stock market.}
\centering
\label{vsHuman}
\begin{tabular}{crrr}
\hline
\multicolumn{1}{|c|}{\begin{tabular}[c]{@{}c@{}}North\\ America\end{tabular}} & \multicolumn{1}{c|}{RIC-NN}     & \multicolumn{1}{c|}{\begin{tabular}[c]{@{}c@{}}RIC-NN\\ (After cost deduction)\end{tabular}}     & \multicolumn{1}{c|}{Funds}   \\ \hline
\multicolumn{1}{|c|}{AR}                                                      & \multicolumn{1}{r|}{9.09\%}     & \multicolumn{1}{r|}{7.79\%}                                                       & \multicolumn{1}{r|}{5.90\%}  \\ \hline
\multicolumn{1}{|c|}{RISK}                                                    & \multicolumn{1}{r|}{17.78\%}    & \multicolumn{1}{r|}{17.78\%}                                                      & \multicolumn{1}{r|}{14.91\%} \\ \hline
\multicolumn{1}{|c|}{R/R}                                                     & \multicolumn{1}{r|}{0.51}       & \multicolumn{1}{r|}{0.44}                                                         & \multicolumn{1}{r|}{0.40}    \\ \hline \hline
\multicolumn{1}{|c|}{\begin{tabular}[c]{@{}c@{}}Asia\\ Pacific\end{tabular}} & \multicolumn{1}{c|}{RIC-NN (TF)} & \multicolumn{1}{c|}{\begin{tabular}[c]{@{}c@{}}RIC-NN (TF)\\ (After cost deduction)\end{tabular}} & \multicolumn{1}{c|}{Funds}   \\ \hline
\multicolumn{1}{|c|}{AR}                                                      & \multicolumn{1}{r|}{12.08\%}    & \multicolumn{1}{r|}{9.44\%}                                                       & \multicolumn{1}{r|}{7.88\%}  \\ \hline
\multicolumn{1}{|c|}{RISK}                                                    & \multicolumn{1}{r|}{17.23\%}    & \multicolumn{1}{r|}{17.23\%}                                                      & \multicolumn{1}{r|}{17.58\%} \\ \hline
\multicolumn{1}{|c|}{R/R}                                                     & \multicolumn{1}{r|}{0.70}       & \multicolumn{1}{r|}{0.55}                                                         & \multicolumn{1}{r|}{0.45}    \\ \hline
\end{tabular}
\end{table}

\section{Conclusion}

In this paper, we have proposed a new stock price prediction framework called RIC-NN by introducing three novel ideas: (1) a nonlinear multi-factor approach, (2) a stopping criteria based on rank IC and (3) deep transfer learning.

RIC-NN is conceptually simple yet universal: The identical NN architecture and RankIC stopping value yielded a consistently good return for a long timescale and the two different markets of very different structures. Experimental comparison showed that RIC-NN outperforms off-the-shell machine learning methods and average performance of investment funds in the last decades.

Directions of promising future work includes the followings.

\textbf{More sophisticated portfolio strategies:} 
In this study, we use a simple equally-weighted (EW) portfolio that maximizing the predictive power of stock returns. 
On the other hand, the portfolio theory \cite{markowitz1952} states that explicit consideration of the risk in portfolio selection is important.
Regarding this direction, combining our method with more sophisticated portfolio strategies, such as Subset Resampling Portfolio \cite{Shen:2017:PSV:3298239.3298460} or Ensemble Growth Optimal Portfolio \cite{shen2019} will be an interesting direction for the future work.

\textbf{Stateful models:} This paper considered a rolling-horizon learning of a neural network, whereas there are several other approaches for portfolio selection. 
In particular, the recurrent neural networks and its variants are stateful neural networks that can capture the time evolution of the stock universe. 
Note that our RIC-NN model uses quite a large time window (i.e., ten years) for the training, which implies that the long-range interaction is important in the multi-factor machine learning models. 
While we presume that a straightforward application of recurrent neural network overfits to the data up to the current time horizon, several attempts to capture long-range interactions, such as memory networks and attention mechanisms, can be applied to predict cross-sectional investments.


\appendix
\section{Appendix A: List of Factors Used in This Paper}
Table \ref{factorlist} shows the list of 20 factors used in this paper.
The financial data is acquired from Compustat, WorldScope and Reuters Fundamentals (ordered by the priority). 
Note that the Compustat data source, which is mainly used to build factors for North America, involves a delay of maximum three months, and the other sources, which are mainly used for Asia Pacific, involve a delay of four months. 
These data sources are used to calculate the factors from No. 1 to No. 14.
The Earnings per share (EPS) revisions, which indicate the future value of a company, are obtained from Thomson Reuters Estimates and I/B/E/S Estimates (ordered by the priority).

We can classify these factors into three types; technical, fundamental and both.
Technical factors are calculated from historical stock prices, whereas fundamental factors are calculated from the qualitative and quantitative information of a company \cite{cavalcante2016}.

\begin{table*}[b!]
\centering
\caption{List of the factors. F (resp. T) in the ``Type'' column indicates that the corresponding factor is derived from its fundamental (resp. technical) property of the stock, and B in the column indicates the factor derived from both of the fundamental and technical properties.}
\begin{tabular}{|c|l|l|l|}
\hline
No & \multicolumn{1}{c|}{Factor} & \multicolumn{1}{c|}{Description}  & \multicolumn{1}{c|}{Type}           \\ \hline
1  & Book-value to Price Ratio                           & Net Asset/Market Value & B                           \\ \hline
2  & Earnings to Price Ratio                           & Net Profit/Market Value & B                      \\ \hline
3  & Dividend Yield                           & Dividend/Market Value & B                            \\ \hline
4  & Sales to Price Ratio                           & Sales/Market Value & B                         \\ \hline
5  & Cash Flow to Price Ratio                          & 
Operating cash flow/Market Value    & B              \\ \hline
6  & Return on Equity                           & Net Profit/Net Asset & F                          \\ \hline
7  & Return on Asset                           &  Net Operating Profit/Total Asset  & F \\ \hline
8  & Return on Invested Capital                          & Net Operating Profit After Taxes/(Liabilities with interest + Net Asset) & F \\ \hline
9  & Accruals                       & -(Changes in Current Assets and Liability-Depreciation)/Total Asset & F \\ \hline
10 & Total Asset Growth Rate                        & Change Rate of Total Assets from the previous period  & F \\ \hline
11 & Current Ratio                          & Current Asset/Current Liability  & F \\ \hline
12 & Equity Ratio                        & Net Asset/Total Asset & F\\ \hline
13 & Total Asset Turnover Rate                        & Sales/Total Asset  & F \\ \hline
14 & Capital Expenditure Growth Rate                      & Change Rate of Capital Expenditure from the previous period & F \\ \hline
15 & EPS Revision (1 month)                & 1 month Earnings Per Share (EPS) Revision & B \\ \hline
16 & EPS Revision (3 month)                & 3 month Earnings Per Share (EPS) Revision & B \\ \hline
17 & Momentum (1 month)                      & Stock Returns in the last month & T\\ \hline
18 & Momentum (12-1 month)            & Stock Returns in the past 12 months except for last month & T \\ \hline
19 & Volatility                       & Standard Deviation of Stock Returns in the past 60 months & T \\ \hline
20 & Skewness                            & Skewness of Stock Returns in the past 60 months & T \\ \hline
\end{tabular}
\label{factorlist}
\end{table*}%

\clearpage 

\section{Appendix B: Additional Experiments}

\subsection{Different Hyperparameters in Off-the-shelf Models}


Tables \ref{tbl:RF_NA} and \ref{tbl:RF_PF} show the results of RF with different depths, whereas Tables \ref{tbl:LASSO_NA} and \ref{tbl:LASSO_PF} show the results of LASSO with different magnitudes of the regularizer. 
Overall, comparison with Tables \ref{tbl:NA} and \ref{tbl:PF} show that (i) RF falls below the competitors, and (ii) LASSO, which outperforms RF, still fall below RIC-NN. 
We have also confirmed that a LASSO with a regularizer stronger than $0.001$ suppresses most of the features, which yields meaningless results.

\begin{table}[htbp]
  \begin{center}
    \begin{tabular}{c}

      \begin{minipage}{0.5\hsize}
\centering
\caption{Results of LASSO with different magnitudes of regularizer in MSCI North America.}
\label{tbl:LASSO_NA}
\begin{tabular}{crrr}
\hline
\multicolumn{1}{|c|}{\multirow{2}{*}{Long}}       & \multicolumn{3}{c|}{Regularizer}                                                              \\ \cline{2-4} 
\multicolumn{1}{|c|}{}                            & \multicolumn{1}{c|}{0.001}    & \multicolumn{1}{c|}{0.0001}   & \multicolumn{1}{c|}{0.00001}  \\ \hline
\multicolumn{1}{|c|}{Alpha}                       & \multicolumn{1}{r|}{0.62\%}   & \multicolumn{1}{r|}{0.53\%}   & \multicolumn{1}{r|}{1.14\%}   \\ \hline
\multicolumn{1}{|c|}{TE}                          & \multicolumn{1}{r|}{5.40\%}   & \multicolumn{1}{r|}{4.60\%}   & \multicolumn{1}{r|}{4.21\%}   \\ \hline
\multicolumn{1}{|c|}{IR}                          & \multicolumn{1}{r|}{0.11}     & \multicolumn{1}{r|}{0.11}     & \multicolumn{1}{r|}{0.27}     \\ \hline
\multicolumn{1}{|c|}{MaxDD}                       & \multicolumn{1}{r|}{-21.84\%} & \multicolumn{1}{r|}{-17.85\%} & \multicolumn{1}{r|}{-16.02\%} \\ \hline \hline
\multicolumn{1}{|l|}{\multirow{2}{*}{Long-Short}} & \multicolumn{3}{c|}{Regularizer}                                                              \\ \cline{2-4} 
\multicolumn{1}{|l|}{}                            & \multicolumn{1}{c|}{0.001}    & \multicolumn{1}{c|}{0.0001}   & \multicolumn{1}{c|}{0.00001}  \\ \hline
\multicolumn{1}{|c|}{AR}                          & \multicolumn{1}{r|}{2.24\%}   & \multicolumn{1}{r|}{0.76\%}   & \multicolumn{1}{r|}{2.00\%}   \\ \hline
\multicolumn{1}{|c|}{RISK}                        & \multicolumn{1}{r|}{10.90\%}  & \multicolumn{1}{r|}{9.56\%}   & \multicolumn{1}{r|}{8.81\%}   \\ \hline
\multicolumn{1}{|c|}{R/R}                         & \multicolumn{1}{r|}{0.21}     & \multicolumn{1}{r|}{0.08}     & \multicolumn{1}{r|}{0.23}     \\ \hline
\multicolumn{1}{|c|}{MaxDD}                       & \multicolumn{1}{r|}{-34.73\%} & \multicolumn{1}{r|}{-35.10\%} & \multicolumn{1}{r|}{-34.52\%} \\ \hline
\end{tabular}
      \end{minipage}

      \begin{minipage}{0.5\hsize}
\centering
\caption{Results of LASSO with different magnitudes of regularizer in MSCI Pacific.}
\label{tbl:LASSO_PF}
\begin{tabular}{crrr}
\hline
\multicolumn{1}{|c|}{\multirow{2}{*}{Long}}       & \multicolumn{3}{c|}{Regularizer}                                                              \\ \cline{2-4} 
\multicolumn{1}{|c|}{}                            & \multicolumn{1}{c|}{0.001}    & \multicolumn{1}{c|}{0.0001}   & \multicolumn{1}{c|}{0.00001}  \\ \hline
\multicolumn{1}{|c|}{Alpha}                       & \multicolumn{1}{r|}{5.35\%}   & \multicolumn{1}{r|}{5.23\%}   & \multicolumn{1}{r|}{4.90\%}   \\ \hline
\multicolumn{1}{|c|}{TE}                          & \multicolumn{1}{r|}{5.17\%}   & \multicolumn{1}{r|}{4.58\%}   & \multicolumn{1}{r|}{4.20\%}   \\ \hline
\multicolumn{1}{|c|}{IR}                          & \multicolumn{1}{r|}{1.04}     & \multicolumn{1}{r|}{1.14}     & \multicolumn{1}{r|}{1.17}     \\ \hline
\multicolumn{1}{|c|}{MaxDD}                       & \multicolumn{1}{r|}{-11.53\%} & \multicolumn{1}{r|}{-8.43\%}  & \multicolumn{1}{r|}{-6.81\%}  \\ \hline \hline
\multicolumn{1}{|l|}{\multirow{2}{*}{Long-Short}} & \multicolumn{3}{c|}{Regularizer}                                                              \\ \cline{2-4} 
\multicolumn{1}{|l|}{}                            & \multicolumn{1}{c|}{0.001}    & \multicolumn{1}{c|}{0.0001}   & \multicolumn{1}{c|}{0.00001}  \\ \hline
\multicolumn{1}{|c|}{AR}                          & \multicolumn{1}{r|}{10.27\%}  & \multicolumn{1}{r|}{10.42\%}  & \multicolumn{1}{r|}{8.99\%}   \\ \hline
\multicolumn{1}{|c|}{RISK}                        & \multicolumn{1}{r|}{9.23\%}   & \multicolumn{1}{r|}{8.35\%}   & \multicolumn{1}{r|}{8.08\%}   \\ \hline
\multicolumn{1}{|c|}{R/R}                         & \multicolumn{1}{r|}{1.11}     & \multicolumn{1}{r|}{1.25}     & \multicolumn{1}{r|}{1.11}     \\ \hline
\multicolumn{1}{|c|}{MaxDD}                       & \multicolumn{1}{r|}{-18.07\%} & \multicolumn{1}{r|}{-13.30\%} & \multicolumn{1}{r|}{-13.34\%} \\ \hline
\end{tabular}
      \end{minipage}

    \end{tabular}
  \end{center}
\end{table}

\begin{table}[htbp]
  \begin{center}
    \begin{tabular}{c}

      \begin{minipage}{0.5\hsize}
\centering
\caption{Results of RF with different depths in MSCI North America.}
\label{tbl:RF_NA}
\begin{tabular}{lrrr}
\hline
\multicolumn{1}{|c|}{\multirow{2}{*}{Long}}       & \multicolumn{3}{c|}{Depth}                                                                    \\ \cline{2-4} 
\multicolumn{1}{|c|}{}                            & \multicolumn{1}{c|}{3}        & \multicolumn{1}{c|}{5}        & \multicolumn{1}{c|}{7}        \\ \hline
\multicolumn{1}{|c|}{Alpha}                       & \multicolumn{1}{r|}{0.77\%}   & \multicolumn{1}{r|}{0.85\%}   & \multicolumn{1}{r|}{0.92\%}   \\ \hline
\multicolumn{1}{|c|}{TE}                          & \multicolumn{1}{r|}{4.80\%}   & \multicolumn{1}{r|}{4.61\%}   & \multicolumn{1}{r|}{4.33\%}   \\ \hline
\multicolumn{1}{|c|}{IR}                          & \multicolumn{1}{r|}{0.16}     & \multicolumn{1}{r|}{0.18}     & \multicolumn{1}{r|}{0.21}     \\ \hline
\multicolumn{1}{|c|}{MaxDD}                       & \multicolumn{1}{r|}{-21.76\%} & \multicolumn{1}{r|}{-23.21\%} & \multicolumn{1}{r|}{-21.16\%} \\ \hline \hline
\multicolumn{1}{|l|}{\multirow{2}{*}{Long-Short}} & \multicolumn{3}{c|}{Depth}                                                                    \\ \cline{2-4} 
\multicolumn{1}{|l|}{}                            & \multicolumn{1}{c|}{3}        & \multicolumn{1}{c|}{5}        & \multicolumn{1}{c|}{7}        \\ \hline
\multicolumn{1}{|c|}{AR}                          & \multicolumn{1}{r|}{1.62\%}   & \multicolumn{1}{r|}{2.44\%}   & \multicolumn{1}{r|}{2.38\%}   \\ \hline
\multicolumn{1}{|c|}{RISK}                        & \multicolumn{1}{r|}{11.10\%}  & \multicolumn{1}{r|}{10.47\%}  & \multicolumn{1}{r|}{9.86\%}   \\ \hline
\multicolumn{1}{|c|}{R/R}                         & \multicolumn{1}{r|}{0.15}     & \multicolumn{1}{r|}{0.23}     & \multicolumn{1}{r|}{0.24}     \\ \hline
\multicolumn{1}{|c|}{MaxDD}                       & \multicolumn{1}{r|}{-40.78\%} & \multicolumn{1}{r|}{-34.90\%} & \multicolumn{1}{r|}{-34.10\%} \\ \hline
\end{tabular}
      \end{minipage}

      \begin{minipage}{0.5\hsize}
\centering
\caption{Results of RF with different depths in MSCI Pacific.}
\label{tbl:RF_PF}
\begin{tabular}{lrrr}
\hline
\multicolumn{1}{|c|}{\multirow{2}{*}{Long}}       & \multicolumn{3}{c|}{Depth}                                                                    \\ \cline{2-4} 
\multicolumn{1}{|c|}{}                            & \multicolumn{1}{c|}{3}        & \multicolumn{1}{c|}{5}        & \multicolumn{1}{c|}{7}        \\ \hline
\multicolumn{1}{|c|}{Alpha}                       & \multicolumn{1}{r|}{2.48\%}   & \multicolumn{1}{r|}{3.53\%}   & \multicolumn{1}{r|}{4.24\%}   \\ \hline
\multicolumn{1}{|c|}{TE}                          & \multicolumn{1}{r|}{5.28\%}   & \multicolumn{1}{r|}{4.96\%}   & \multicolumn{1}{r|}{4.94\%}   \\ \hline
\multicolumn{1}{|c|}{IR}                          & \multicolumn{1}{r|}{0.47}     & \multicolumn{1}{r|}{0.71}     & \multicolumn{1}{r|}{0.86}     \\ \hline
\multicolumn{1}{|c|}{MaxDD}                       & \multicolumn{1}{r|}{-13.04\%} & \multicolumn{1}{r|}{-9.21\%}  & \multicolumn{1}{r|}{-7.36\%}  \\ \hline \hline
\multicolumn{1}{|l|}{\multirow{2}{*}{Long-Short}} & \multicolumn{3}{c|}{Depth}                                                                    \\ \cline{2-4} 
\multicolumn{1}{|l|}{}                            & \multicolumn{1}{c|}{3}        & \multicolumn{1}{c|}{5}        & \multicolumn{1}{c|}{7}        \\ \hline
\multicolumn{1}{|c|}{AR}                          & \multicolumn{1}{r|}{6.48\%}   & \multicolumn{1}{r|}{7.83\%}   & \multicolumn{1}{r|}{8.74\%}   \\ \hline
\multicolumn{1}{|c|}{RISK}                        & \multicolumn{1}{r|}{8.63\%}   & \multicolumn{1}{r|}{8.61\%}   & \multicolumn{1}{r|}{8.87\%}   \\ \hline
\multicolumn{1}{|c|}{R/R}                         & \multicolumn{1}{r|}{0.75}     & \multicolumn{1}{r|}{0.91}     & \multicolumn{1}{r|}{0.99}     \\ \hline
\multicolumn{1}{|c|}{MaxDD}                       & \multicolumn{1}{r|}{-20.09\%} & \multicolumn{1}{r|}{-15.13\%} & \multicolumn{1}{r|}{-13.56\%} \\ \hline
\end{tabular}
      \end{minipage}

    \end{tabular}
  \end{center}
\end{table}
\clearpage

\section{Appendix C: Cumulative Return of RIC-NN and Investment Funds}

\begin{figure}[b!]
\centering
 \begin{minipage}{\hsize}
  \begin{center}
   \includegraphics[width=0.8\hsize]{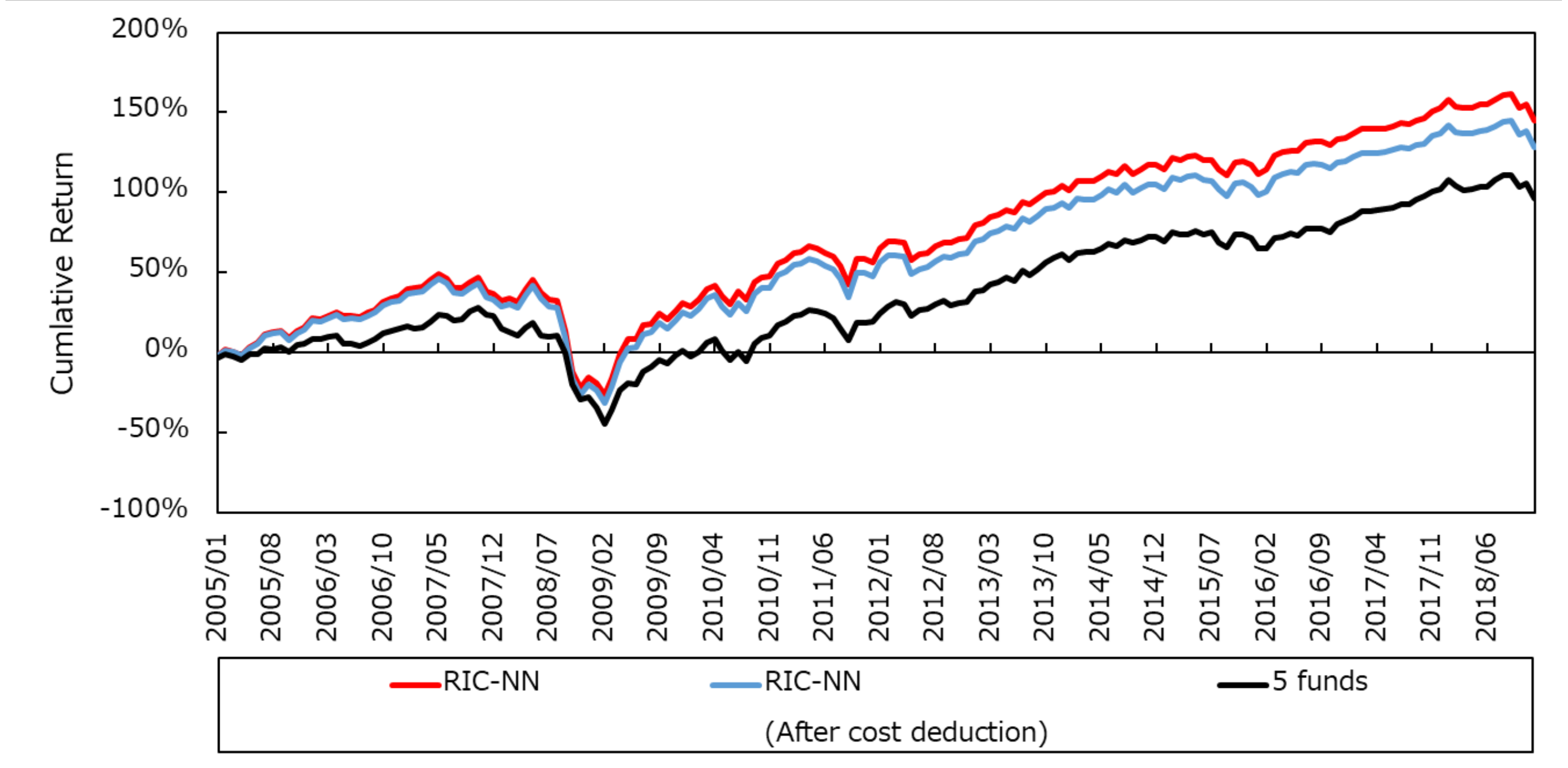}
  \end{center}
  \caption{Averaged cumulative returns among the five funds and the cumulative return of RIC-NN in MSCI North America. RIC-NN consistently outperforms the average of the funds. Both of the averaged funds and RIC-NN suffer a large drawdown during the financial crisis in 2007-2008.} 
  \label{fig:HumanNA_Return}
 \end{minipage}
 \begin{minipage}{\hsize}
  \begin{center}
   \includegraphics[width=0.8\hsize]{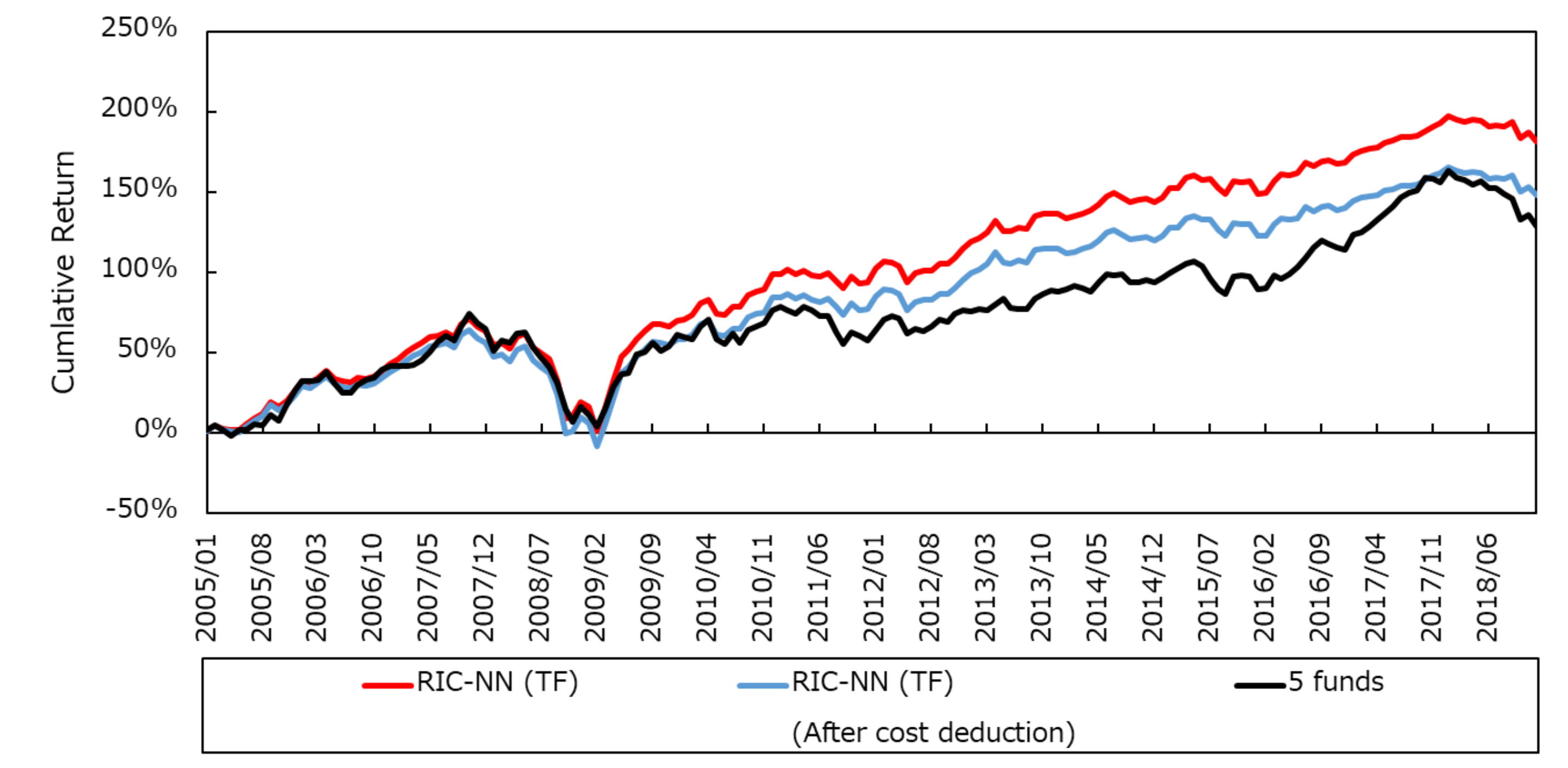}
  \end{center}
  \caption{Averaged cumulative returns among the five funds and the cumulative return of RIC-NN in MSCI Pacific. }
  \label{fig:HumanPF_Return}
 \end{minipage}
\end{figure}

Figure \ref{fig:HumanNA_Return} and \ref{fig:HumanPF_Return} show the  corresponding return time series in the NA and AP regions, respecively. Regarding the evaluated risk and returns, see Section ``Comparison with Actual Investment Funds'' in the main paper.

\end{document}